\documentclass[sigconf]{acmart}
\usepackage{booktabs} 
\usepackage{amsmath,url,graphicx,amssymb}
\usepackage{amsfonts}
\usepackage{ctable}
\usepackage{multirow}
\usepackage{algorithm}
\usepackage{algpseudocode}
\usepackage{pifont}
\usepackage{color}
\usepackage{subfigure}
\usepackage{enumitem}
\usepackage{sidecap}

\newcommand{\rom}[1]{\uppercase\expandafter{\romannumeral #1\relax}}

\newcommand{\mylistbegin}{
  \begin{list}{$\bullet$}
   {
     \setlength{\itemsep}{-2pt}
     \setlength{\leftmargin}{1em}
     \setlength{\labelwidth}{1em}
     \setlength{\labelsep}{0.5em} } }
\newcommand{\mylistend}{
   \end{list}  }

\newcommand{\eg}{\textit{e.g.}}

\newcommand{\ie}{\textit{i.e.}}
\newcommand{\etc}{\textit{etc}}

\newcommand{\wrt}{\textit{w.r.t.~}}

\newcommand{\header}[1]{{\vspace{+1mm}\flushleft \textbf{#1}}}

\newtheorem{thm:def}{Definition}

\newcommand{\mcW}{\mathcal{W}}
\newcommand{\mcY}{\mathcal{Y}}

\newcommand{\mcA}{\mathcal{A}}
\newcommand{\mcV}{\mathcal{V}}
\newcommand{\mcE}{\mathcal{E}}
\newcommand{\mcG}{\mathcal{G}}
\newcommand{\mcX}{\mathcal{X}}
\newcommand{\mcD}{\mathcal{D}}
\newcommand{\mcZ}{\mathcal{Z}}
\newcommand{\mcH}{\mathcal{H}}
\newcommand{\mcN}{\mathcal{N}}
\newcommand{\mcC}{\mathcal{C}}
\newcommand{\mcU}{\mathcal{U}}

\setcopyright{rightsretained}
\acmDOI{10.475/123_4}
\acmISBN{123-4567-24-567/08/06}

\acmConference[]{}{}{} 
\acmYear{2018}
\copyrightyear{2018}

\acmPrice{15.00}

\begin{document}
\title{Relation Learning on Social Networks with \mbox{Multi-Modal Graph Edge Variational Autoencoders}}
\author{\Large{Carl Yang$^*$, Jieyu Zhang$^*$, Haonan Wang$^*$, Sha Li$^*$, Myungwan Kim$^\#$, Matt Walker$^\#$, Yiou Xiao$^\#$, Jiawei Han$^*$}}
       \affiliation{
       \institution{$^*$University of Illinois, Urbana Champaign, 201 N. Goodwin Ave, Urbana, IL 61801, USA}
       \institution{$^\#$LinkedIn Co., 599 N. Mathilda Ave, Sunnyvale, CA 94085, USA}
       \institution{$^*$\{jiyang3, jieyuz2, haonan3, shal2, hanj\}@illinois.edu, $^\#$\{mukim, mtwalker, yixiao\}@linkedin.com}
       }

\setlength{\floatsep}{4pt plus 4pt minus 1pt}
\setlength{\textfloatsep}{4pt plus 2pt minus 2pt}
\setlength{\intextsep}{4pt plus 2pt minus 2pt}

\setlength{\dbltextfloatsep}{3pt plus 2pt minus 1pt}
\setlength{\dblfloatsep}{3pt plus 2pt minus 1pt}
\setlength{\abovecaptionskip}{3pt}
\setlength{\belowcaptionskip}{2pt}
\setlength{\abovedisplayskip}{2pt plus 1pt minus 1pt}
\setlength{\belowdisplayskip}{2pt plus 1pt minus 1pt}
\renewcommand\footnotetextcopyrightpermission[1]{}
\pagenumbering{gobble}
\renewcommand{\shortauthors}{Carl Yang \textit{et al.}}
\settopmatter{printacmref=false, printfolios=false}

\begin{abstract}
While node semantics have been extensively explored in social networks, little research attention has been paid to profile edge semantics, \ie, social relations. 
Ideal edge semantics should not only show that two users are connected, but also why they know each other and what they share in common. 
However, relations in social networks are often hard to profile, due to noisy multi-modal signals and limited user-generated ground-truth labels.

In this work, we aim to develop a unified and principled framework that can profile user relations as edge semantics in social networks by integrating multi-modal signals in the presence of noisy and incomplete data.
Our framework is also flexible towards limited or missing supervision.
Specifically, we assume a latent distribution of multiple relations underlying each user link, and learn them with multi-modal graph edge variational autoencoders. We encode the network data with a graph convolutional network, and decode arbitrary signals with multiple reconstruction networks.
Extensive experiments and case studies on two public DBLP author networks and two internal LinkedIn member networks demonstrate the superior effectiveness and efficiency of our proposed model.
\end{abstract}

\maketitle
\section{Introduction}
\label{sec:intro}
On social networks, while nodes are explicitly associated with rich contents (\eg, attributes, diffusions), the \textit{semantics} of each link is often implicit. Without such semantics, we cannot truly understand the interaction between users.
In this work, we propose and study the problem of \textit{relation learning on social networks}. The goal is to learn the relation semantics underlying each existing link in the social network, which naturally improves the \textit{targeting} of various downstream services, such as friend suggestion, attribute profiling, user clustering, influence maximization and recommendation.

Unlike relation prediction or extraction among entities \cite{bordes2013translating, gardner2015efficient, socher2013reasoning, wang2015knowledge, wang2014knowledge, jiang2017metapad, lin2016neural, liu2017heterogeneous, wang2015constrained, zeng2014relation}, relation learning on social networks is hard due to the anonymous nature of users, lack of large-scale free-text as context, and very limited labeled data \cite{yang2019relationship}. Moreover, information on social networks is \textit{multi-modal}, \textit{noisy} and \textit{incomplete} \cite{yang2019cubenet, yang2018know}, leading to various useful but low-quality signals, which are challenging for a unified model to properly \textit{regulate} and \textit{integrate}.

\begin{figure*}[t!]
\centering
\subfigure[Latent relations]{
\includegraphics[width=0.242\textwidth]{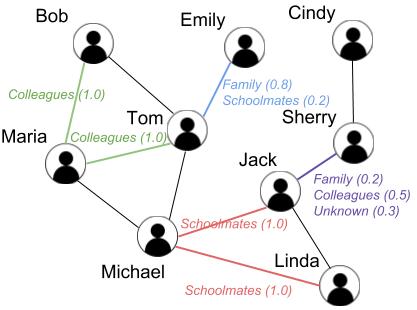}}
\subfigure[Leveraging network proximities]{
\includegraphics[width=0.242\textwidth]{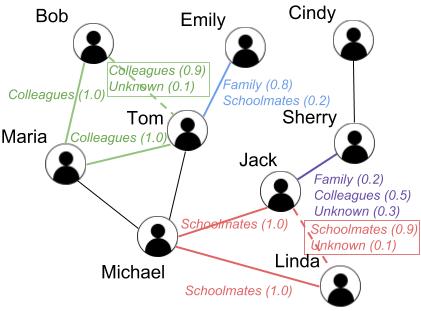}}
\subfigure[Leveraging user attributes]{
\includegraphics[width=0.242\textwidth]{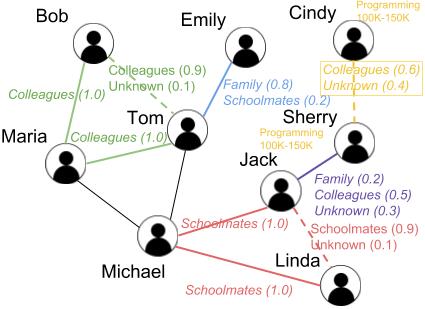}}
\subfigure[Leveraging information diffusions]{
\includegraphics[width=0.242\textwidth]{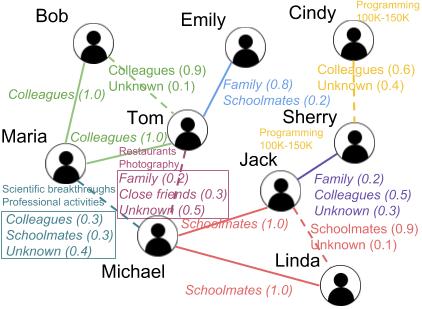}}
\caption{\textbf{A running toy example of a LinkedIn social network of 9 users and 11 links.}}
\label{fig:toy}
\end{figure*}

Figure \ref{fig:toy} gives an example of a toy social network. 
As shown in (a), we assume the existence of some latent relation(s) for each link in the network. 
For example, \textsf{Tom} and \textsf{Maria} are \textsf{colleagues}, whereas \textsf{Jack} and \textsf{Michael} are \textsf{schoolmates}. 
Furthermore, to better reflect reality, we model each link with a \textit{relation distribution}. For example, the relationship between \textsf{Tom} and \textsf{Emily} is built up by $80\%$ \textsf{relatives} and $20\%$ \textsf{schoolmates}, \ie, they are from the same family, which makes the \textsf{relative} relation dominate their link, but they also go to the same school, thus forming a weaker \textsf{schoolmate} relation. 
We also allow a link to carry an \textsf{unknown} relation, modeling the uncertainty of relation strength.

This example also demonstrates three types of signals that are helpful in relation inference. 

\begin{itemize}[leftmargin=10pt]
\item \textbf{Network proximity.} As illustrated in Figure \ref{fig:toy} (b), the network structure is highly useful for inferring unknown relations. If we are confident that \textsf{Tom} and \textsf{Maria} as well as \textsf{Maria} and \textsf{Bob} are \textsf{colleagues}, we can easily deduce that \textsf{Tom} and \textsf{Bob} are also \textsf{colleagues}. Similar situations exist for other pairs like \textsf{Jack} and \textsf{Linda}, who are likely \textsf{schoolmates}.

\item \textbf{User attribute.} As the \textit{homophily} phenomenon~\cite{mcpherson2001birds, yang2017bi} suggests, user attributes can be highly indicative of their relations. As shown in Figure \ref{fig:toy} (c), if \textsf{Cindy} and \textsf{Sherry} share similar skills (\textsf{programming}) and salary level (\textsf{100K-150K}), their relation is more likely to be \textsf{colleagues} (\eg, $60\%$) than others (\eg, $40\%$).

\item \textbf{Information diffusion.} As shown in Figure \ref{fig:toy} (d), users on social networks often interact in different ways, where links become biased information routes. For example, \textsf{Maria} often shares \textsf{Michael}'s posts about \textsf{scientific breakthroughs} or \textsf{professional activities}, while \textsf{Tom} likes to comment on \textsf{Michael}'s posts on \textsf{restaurants} and \textsf{photography}. Intuitively, it is more likely for \textsf{Maria} and \textsf{Michael} to be \textsf{colleagues} or \textsf{schoolmates}, and \textsf{Tom} and \textsf{Michael} to be \textsf{relatives} or \textsf{close friends}.
\end{itemize}

Note that in real-world social networks, each of the three types of information can be highly noisy and incomplete. 
Moreover, high-quality training data is highly limited, if any. This requires a model for relation learning on social networks to be:
(1) \textit{powerful} to fully leverage and coherently integrate the multi-modal signals;
(2) \textit{robust} to produce reliable results when certain data are missing or inaccurate;
(3) \textit{flexible} to operate with limited or no supervision.

In the face of such challenges, we develop \textsc{ReLearn}, a unified multi-modal graph edge variational autoencoder framework. 
Essentially, our model belongs to the class of unsupervised representation learning models using autoencoders, which has been shown effective for various machine learning tasks \cite{schmidhuber2015deep, vincent2008extracting, le2013building}.
On top of it, we design a Gaussian mixture variational autoencoder to encode link semantics, with the mixture weights representing the distribution over relation types \textit{local} to the link. 
We further assume \textit{global} relation prototype variables for the latent relations, which are instantiated as a Gaussian distribution in our model.
Variational inference with two-step Monte Carlo sampling is designed to infer both the global Gaussian parameters and local relation distributions.

To compute graph edge representations on large-scale social networks, we combine graph convolutional networks (GCN) \cite{kipf2016semi} with fully-connected feedforward networks (FNN) for our encoder, and enable batch-wise training with fixed-size neighborhood sampling.
To fully leverage and integrate multi-modal signals, we attach multiple decoders to the GCN-based encoder,
which can be flexibly trained with any combination of available signals.
Finally, the framework can be trained with varying amount of labeled data by using the labels as priors in the objective function. 

We conduct extensive experiments on four real-world large-scale social networks, \ie, two public DBLP author networks and two internal LinkedIn member networks\footnote{DBLP source: https://dblp.uni-trier.de/; LinkedIn source: https://www.linkedin.com/}. Through the comparison with various state-of-the-art baselines, we observe consistent significant improvements of $8$\%-$28$\% over the best baselines. The generative nature of \textsc{ReLearn} further enables interpretable case studies that provide insights into the learned relations. 

The main contributions of this work are summarized as follows. 
\begin{itemize}[leftmargin=10pt]
\item We propose and formulate the problem of relation learning on social networks as finding the hidden semantics underlying user links, and study its implication towards various applications.
\item We develop \textsc{ReLearn}, a powerful, robust and flexible relation learning framework by leveraging social network signals including network proximity, user attributes and information diffusion.
\item We conduct comprehensive experiments on four real-world social networks with different model variants to demonstrate the effectiveness of the proposed techniques.
\end{itemize}

\section{Related Work and Preliminaries}
\label{sec:related}
\subsection{Social Network Analysis}
Some works on social network analysis have looked into the latent relations underlying uniform social links. Among them, \cite{chakrabarti2014joint, he2007graph, tang2009relational} aim to jointly learn user attributes and relations, by assuming the relations to be mutually exclusive and determined by user attributes, whereas \cite{leskovec2012learning, yang2013community, yang2009combining, ruan2013efficient} attempt to detect groups constructed by homogeneous relations. While both groups of methods implicitly learn the relation semantics, their assumptions about relations are restricted and unrealistic, since relations are not necessarily mutually exclusive and are not only learnable among groups. Moreover, their methods also do not integrate various signals as we consider in this work.  \cite{tu2017cane} leverages text context to encode relation semantics in node embeddings. In comparison, we directly learn edge representations and text is only one of the signals we consider.

\subsection{Relation Learning in Other Contexts}
The problem of relation learning has been intensively studied in knowledge graph completion and relation extraction. Some existing works rely more on the reasoning over existing knowledge graphs with typed links \cite{bordes2013translating, gardner2015efficient, socher2013reasoning, wang2015knowledge, wang2014knowledge}, while others leverage more on the modeling of textual contexts with weak supervision \cite{jiang2017metapad, lin2016neural, liu2017heterogeneous, wang2015constrained, zeng2014relation}. However, on social networks, nodes are untyped as well as links, and they are often anonymous without textual contexts. On the other hand, noisy signals like link structures, user attributes and information diffusions widely exist, which urges us to develop novel models for relation learning on social networks.

\subsection{Related Techniques}
\subsubsection{Network Embedding}
After the great success of DeepWalk \cite{perozzi2014deepwalk}, network embedding has attracted much research attention in recent years. 
We mainly compare with those on content-rich networks. For example, models like TADW \cite{yang2015network}, PTE \cite{tang2015pte}, Planetoid \cite{yang2016revisiting}, paper2vec \cite{ganguly2017paper2vec}, STNE \cite{liu2018content}, AutoPath \cite{yang2018similarity} and NEP \cite{yang2019neural} have been designed to improve network embedding by incorporating node contents like types, attributes and texts. Moreover, the convolution based models like GCN \cite{kipf2016semi}, GAT \cite{velickovic2017graph}, GraphSage \cite{hamilton2017inductive}, CANE \cite{tu2017cane}, DiffPool \cite{ying2018hierarchical}, JK-Net \cite{xu2018representation}, FastGCN \cite{chen2018fastgcn} and DGI \cite{velivckovic2018deep} naturally take the input of both node features and links. However, most of them cannot be trained in an unsupervised fashion, and none of them can easily incorporate additional signals like information diffusions on networks.

Moreover, a few recent works on diffusion prediction also computes network embedding by modeling the diffusions as DAGs or trees, such as CDSK \cite{bourigault2014learning}, DCB \cite{atwood2016diffusion}, EmbIC \cite{bourigault2016representation}, TopoLSTM \cite{wang2017topological} and inf2vec \cite{feng2018inf2vec}. In this way, they combine the signals of diffusions and network links. However, they often only care about local network embedding that captures the diffusion structures rather than all links on the network, and they do not integrate node contents.

To the best of our knowledge, our model is the first one to seamlessly incorporate various signals for robust graph edge embedding, and is able to work when any of the signals are missing or more additional signals become available.

\subsubsection{Variational Autoencoders}
Variational autoencoders (VAEs) \cite{Kingma2013AutoEncodingVB, rezende2014stochastic} combine Bayesian inference with the flexibility of neural networks for robust representation learning. 
By applying the reparameterization trick, VAE allows the use of standard backpropagation to optimize continuous stochastic variables. 
In its simplest form, VAE can be viewed as a one-layer latent variable model: 
\begin{equation}
p(x,z) = p(z) p(x|z)
\end{equation}
where $x$ is an observed variable and $z$ is a hidden variable.
Using variational inference, the goal is to maximize the evidence lower bound (ELBO):
\begin{equation}
\begin{aligned}
\mathcal { L } \left( p _ { \theta } , q _ { \phi } \right) &= \mathbb { E } _ { q _ { \phi } ( z | x ) } \left[ \log p _ { \theta } ( x , z ) - \log q _ { \phi } ( z | x ) \right] \\
& = \mathbb { E } _ { q _ { \phi } ( z | x ) } \left[ \log p _ { \theta } ( x | z ) \right] - K L \left( q _ { \phi } ( z | x ) \| p ( z ) \right).
\end{aligned}
\end{equation}
We refer readers to \cite{Kingma2013AutoEncodingVB} for the derivation of this lower bound.
Both $q_\phi(z|x)$ and $p_\theta(x|z)$ are parameterized by neural networks. They are referred to as the \textit{encoder network} and the \textit{decoder network}, respectively. The first term in the ELBO is a reconstruction loss that encourages the decoded $x$ to be close to the observed $x$. The second term is a regularization term where the posterior distribution of $z$ is pulled towards the prior, which is often a simple distribution.


To extend the use of VAE to discrete variables, \cite{jang2017categorical, maddison2017concrete} introduced the Gumbel-Softmax distribution which is a continuous approximation of categorical variables. 
Given a categorical variable $z$ and its class probabilities $\pi_1, \ldots, \pi_k$, we can sample from this distribution by first sampling $k$ times from the Gumbel(0,1) distribution. The \textit{argmax} operation in the original Gumbel-Max trick is replaced by a softmax operation to ensure the differentiability of the function
\begin{equation}
z _ { i } = \frac { \exp \left( \left( \log \left( \pi _ { i } \right) + g _ { i } \right) / \tau \right) } { \sum _ { j = 1 } ^ { k } \exp \left( \left( \log \left( \pi _ { j } \right) + g _ { j } \right) / \tau \right) }, \quad \text { for } i = 1 , \ldots , k.
\label{eq:gumbel-softmax}
\end{equation}

As we will show later, the key technical innovation of this work lies in our deliberate design of a powerful, robust and flexible relation learning model based on the principled framework of VAE.
\section{\textsc{ReLearn}}
\label{sec:model}
\subsection{Problem Definition}

\header{Input.}
As we have discussed in Section \ref{sec:intro}, we aim to jointly consider multiple signals on social networks that are indicative of relation semantics.
We use a graph $\mcG=\{\mcV, \mcE, \mcA, \mcD\}$ to model all data we consider in this work.
$\mcV=\{v_i\}_{i=1}^N$ is the set of nodes (users). 
$\mcE=\{e_{ij}\}_{i, j=1}^N$ is the set of edges (links), where $e_{ij}=1$ denotes an existing link between $v_i$ and $v_j$, and $e_{ij}=0$ otherwise. We consider undirected links in this work, while the model can be easily generalized for directed links. 
$\mcA$ is the set of node features (user attributes) associated with $\mcV$, where each $a_{i}\in\mcA$ is a fixed-sized vector of dimension $L$ associated with $v_i$. The exact features encoded in $\mcA$ is dataset-dependent and we refer the reader to Section \ref{sec:exp} for details. 
$\mcD=\{d_s\}_{s=1}^M$ is the set of \textit{diffusion induced networks} generated from the information diffusions over the network, which we formally define as follows.

\begin{thm:def}
Diffusion Induced Network. A network $d_s=\{\mcV_s,$ $\mcE_s, \mcC_s\}$ is a diffusion induced network generated by a piece of information $\xi_s$ that flows on the whole network $\mcN=\{\mcV, \mcE\}$, if $\mcV_s\subset\mcV$ is the set of nodes affected by $\xi_s$, $\mcE_s\subset\mcE$ is the set of edges among $\mcV_s$, and $\mcC_s$ is the contents associated with $\xi_s$.
\end{thm:def}

Taking $\mathcal{G}$ as input, our goal is to compute the following output of edge representations $\mcH$, which in an ideal case should encode the underlying relation semantics we aim to learn from $\mathcal{G}$.

\header{Output.}
We aim to output $\mcH=\{h_{ij}\}_{i,j=1}^N$ as a set of edge representations. Each $h_{ij}\in\mcH$ is a fixed-sized vector learned for edge $e_{ij}$. 

We especially care about the representations of existing links (\ie, $e_{ij}=1$), so as to further understand their underlying relation semantics and make relation predictions through generic classification or clustering algorithms.
The representations of non-existing links (\ie, $e_{ij}=0)$ might also be useful for tasks like typed link prediction but is not the focus of this work.

We now formally define the relation learning problem as follows.

\begin{thm:def}
Relation Learning on Social Networks. Given a social network $\mcG=\{\mcV, \mcE, \mcA, \mcD\}$, learn the edge representation $\mcH$ by integrating the multiple signals from $\mcE$, $\mcA$ and $\mcD$, which captures the relations underlying $\mcE$.
\end{thm:def}

%

\subsection{Model}
In this work, we propose \textsc{ReLearn}, a unified model of multi-modal graph edge variational autoencoder. 
It follows a novel design of a single-encoder-multi-decoder framework, so as to coherently model the multi-modal signals on social networks, and flexibly operate when any of the signals are missing. 
A robust Gaussian mixture model with global Gaussian distributions and local mixture weights is injected to regulate the latent edge embedding space and capture the underlying relation semantics.

\subsubsection{Gaussian Mixture Variational Autoencoder}
Motivated by recent success of autoencoders, our idea is to find latent relations that inherently generate the observed various signals on social networks. Following this insight, we believe that the edge representation $\mcH$, as the \textit{codec} computed via encoding and decoding the observed signals through the autoencoder framework, should reflect the underlying relations and follow a certain relation-specific distribution in the embedding space. 

Particularly, we assume $\mcH$ can be further decomposed into the combination of a relation factor $\mcZ$ and an embedding factor $\mcW$:
\begin{align}
h_{ij} = \sum_{k=1}^K z_{ijk} w_{ijk},
\end{align}
where for each pair of nodes $v_i$ and $v_j$, $w_{ij}$ follows the same set of $K$ independent global multivariate Gaussian distributions, \ie,
\begin{align}
\forall k=1,\ldots, K: \; w_{ijk} \sim \mcN(\mu_k, \sigma^2_k),
\end{align}
and $z_{ij}$ follows a local multinomial distribution, \ie,
\begin{align}
z_{ij} \sim Mul(K, \pi_{ij}), \; \pi_{ij} = (\pi_{ij1}, \ldots, \pi_{ijK}).
\end{align}

The idea behind this design is intuitive: We assume there are $K$ possible latent relations, which is directly modeled by the local relation factor $z_{ij}\in\mathbb{R}^K$. The multinomial distribution is chosen to respect the fact that multiple relations can co-exist on the same link.
The edge representation $h_{ij}$ is then a weighted summation over the global embedding factor $w_{ij}\in\mathbb{R}^{K\times P}$. We use a multivariate Gaussian to model the edge semantics as a probability distribution instead of a deterministic value so that   the uncertainty in the data due to noisy and inaccurate signals can be captured by its variance.

Note that, for any pair of nodes $v_i$ and $v_j$, $w_{ij}$ follows the same $K$ global Gaussian components, which are fixed across all edges, while the mixture assignment is inferred on each edge. Such a design helps us largely reduce the number of parameters to be learned for $\mcH$ and alleviate the problem of data sparsity.

\begin{figure}[t!]
	\centering
	\includegraphics[width=0.9\linewidth]{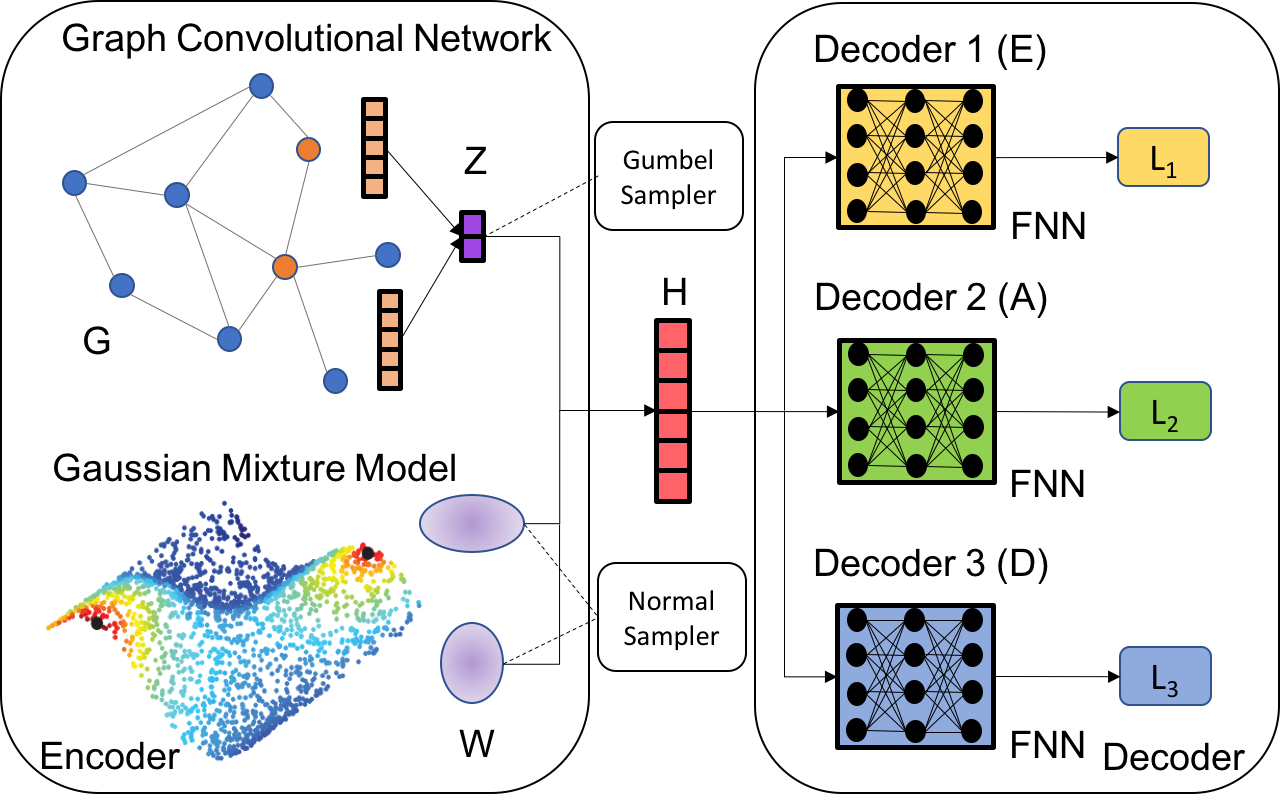}
	\caption{The multi-modal graph edge variational autoencoder architecture of \textsc{ReLearn}.}
	\vspace{-10pt}
	\label{fig:mm}
\end{figure}


\begin{figure}
  \centering
  \includegraphics[width=0.4\linewidth]{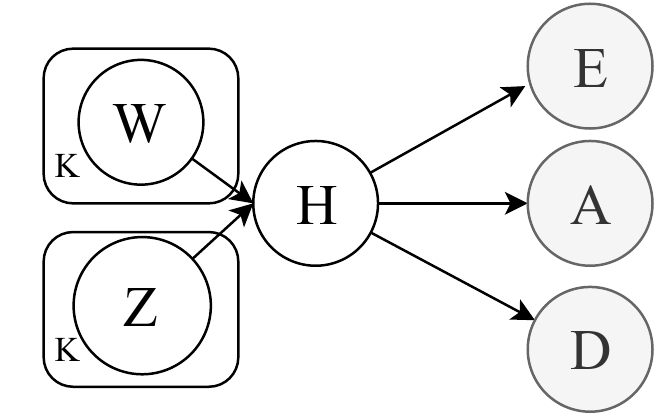}
    \caption{\label{fig:graphic}Plate diagram for our variational autoencoder. $W$ is the embedding factor, $Z$ is the relation factor (mixture weight for the Gaussian random variables), $H$ is the edge embedding, $E$ indicates edge existence, $A$ encodes edge attributes and $D$ encodes diffusion information. 
    All random variables are defined separately for each edge. }
\end{figure}


To learn the edge embedding $\mcH$, we assume that all observable signals on social networks are independently generated given $\mcH$, as reflected in Figure \ref{fig:graphic}.
Consider a particular observed signal $\mcX$ to learn (\eg, if we consider user attribute, then $\mcX=\mcA$), we can derive the corresponding evidence lower bound objective (ELBO):
\begin{equation}
\begin{aligned}
& \mathcal { L } ( p _ { \theta } , q _ { \phi } ) \\
= & \mathbb { E } _ { q _ { \phi } (Z, W,H |X ) } [ \log p _ { \theta } ( Z,W,H, X ) - \log q _ { \phi } ( Z,W,H | X ) ] \\
= & \mathbb { E } _ { q _ { \phi } (Z, W,H |X) }[\log p_\theta (X|H) ]  \\
& - KL[(q_\phi(W) \Vert p(W)] -  KL[q_\phi(Z|X) \Vert p(Z)].  
\label{eq:elbo}
\end{aligned} 
\end{equation}

In the equation, the first term is the reconstruction loss on $\mcX$, which allows the model to extract useful patterns from observed network signals that are indicative of relation semantics. The second and third terms regularize the latent variables towards the priors. When no prior knowledge is available, the unit Gaussian distribution and uniform multinomial distributions can be applied to regularize $\mcW$ and $\mcZ$, respectively. However, when labeled relations are available during training, we can use a smoothed one-hot multinomial distribution per labeled node pair as the prior to effectively inject supervision, \ie,
\begin{align}
p(z_{ij}=k) = \frac{\mathbb{I}(k=z^*_{ij})+\eta}{1+K\eta},
\end{align} 
where $z^*_{ij}$ is the ground-truth relation label on $e_{ij}$ and $\eta$ is a smoothing parameter. 
In this way, our model can flexibly leverage any amount of supervision, and even work under no supervision.


\subsubsection{Graph Edge Encoder}
The goal of our encoder network is to output the local relation factor $\mcZ$, which is combined with the global embedding factor $\mcW$ to generate the edge embedding $\mcH$. 

GCN \cite{kipf2016semi} has been widely used to compute latent representations from node feature and network structure \cite{kipf2016variational, yang2019conditional}. 
To consider multiple signals for edge representations, we design a graph edge encoder based on GCN.
Specifically, we have
\begin{align}
\mcU^{(l+1)} = \text{ReLU}(\tilde{D}^{-\frac{1}{2}}\tilde{E}\tilde{D}^{-\frac{1}{2}}\mcU^{(l)}W_g^{(l)}),
\end{align}
which is a standard GCN layer. In our setting, $\mcU^{(0)}=\mcA$, $\tilde{E}=E+I_N$, $\tilde{D}_{ii}=\sum_j \tilde{E}_{ij}$, and $W_g$ are the learnable GCN parameters. $E$ is the 0-1 edge existence matrix. 
For the sake of scalability, we implement batch-wise training for GCN via fixed-sized neighborhood sampling~\cite{hamilton2017inductive}.

For a pair of nodes $v_i$ and $v_j$ ($i<j$), we concatenate their node features to form an edge feature $\mathbf{y}_{ij}\in\mcY$
\begin{align}
\mathbf{y}_{ij} = [\mathbf{u}_i, \mathbf{u}_j],
\end{align}
where $\mathbf{u}_i, \mathbf{u}_j\in\mcU$ are the node features of $v_i$ and $v_j$, respectively. In this work, we do not differentiate the head and tail nodes for an edge, since we only consider undirected links in the social networks.

Finally, we add a feed-forward neural network (FNN) with ReLU activations that takes edge features to compute the relation factors as $Z = f_r(Y)$. Altogether, the parameters $\phi$ to be learned in the encoder network is $\{\phi_g, \phi_r, \phi_w\}$, where $\phi_g$ is the set of parameters in GCN, $\phi_r$ is the set of parameters in FNN, and $\phi_w$ is the set of parameters in the $K$ global relation-specific Gaussian distributions. Detailed configurations of the GCN and FNN are described in Sec.~\ref{sec:exp}.

\subsubsection{Multi-Modal Decoder}

Figure \ref{fig:mm} illustrates our particular design of multi-modal graph edge variational autoencoder that jointly models the network proximities $\mcE$, user attributes $\mcA$ and information diffusions $\mcD$ on social networks, while various other possibly useful signals can be easily plugged in with flexibility upon availability. 

In this work, the decoder network consists of three decoders, each of which models the generation process of a particular observed signal given the edge representation $\mcH$.
\begin{enumerate}[leftmargin=15pt]
\item A network proximity decoder, which models $p_\theta(\mcE|\mcH)$.
\item A user attribute decoder, which models $p_\theta(\mcA|\mcH)$.
\item An information diffusion decoder, which models $p_\theta(\mcD|\mcH)$.
\end{enumerate}
In Eq.~\ref{eq:elbo}, we used $\mcX$ as a placeholder for any possible signal on $\mcG$. By plugging in all three decoders, we have our final ELBO.
\begin{align}
  & \mathcal { L } ( p _ { \theta } , q _ { \phi } ) = \mathbb{E}_{q_\phi(Z, W, H | G) }[\lambda_1\log p_\theta (E|H) + \lambda_2\log p_\theta (A|H) + \lambda_3 \nonumber\\
  & \log p_\theta (D|H)] - KL[(q_\phi(W) \Vert p(W)] -  KL[q_\phi(Z|G) \Vert p(Z)],
\end{align}
where $\lambda_i$'s are the weighting parameters with $\sum_{i=1}^3\lambda_i=1$.

Each of the three decoders are implemented as simple FNNs. Decoder 1 tries to reconstruct links on the network with the following cross-entropy loss on $\mcE$:
\begin{equation}
\begin{aligned}
L_1=&\mathbb{E}_{q_\phi(Z,W,H|E)} [ \log p_{\theta}(E|H)] =   \sum_{i,j} \mathbb{E}_{h \sim q_\phi} \log p_\theta(e_{ij}|h_{ij}) \\
= & \sum_{i,j} \{ e_{ij}\log \varsigma (f_{d1}(h_{ij})) + (1-e_{ij})\log [1-\varsigma (f_{d1}(h_{ij}))]\}, 
\end{aligned}
\end{equation}
where $\varsigma(x)=\frac{1}{1+e^{-x}}$ is the sigmoid function and $f_{d1}$ is the FNN of decoder 1. During training, we sample positive and negative pairs of nodes, where positive samples are from node pairs with observed links (\ie, $e_{ij}=1$) on $\mcG$, and for each positive pair, we randomly corrupt one end of the link to get negative samples.

Decoder 2 tries to recover the edge attributes, which are the concatenations of node (user) attributes on the two ends (\ie, $a_{ij}=[a_i, a_j]$). It computes an $\ell_{2}$ loss on $\mcA$ (constant terms omitted):
\begin{equation}
\begin{aligned}
L_2=& \mathbb{E}_{q_\phi(Z,W,H|A)}[ \log p_{\theta}(A|H)] \\
= & \sum_{i,j} \mathbb{E}_{h\sim q_\phi} \log p_{\theta}(a_{ij}|h_{ij}) = \sum_{i,j} \mathbb{E}_{h\sim q_\phi} \Vert a_{ij}-f_{d2}(h_{ij}) \Vert_2^2,
\end{aligned}
\end{equation} 
where $f_{d2}$ is the FNN of decoder 2.  Since $h_{ij}$ is the generated from the two-step Monte Carlo sampling, variance has been pushed to the encoder parameters $\mcZ$ and $\mcW$. 

Decoder 3 tries to recover diffusion contents on links covered by the corresponding diffusions, by computing a similar $\ell_{2}$ loss on $\mcD$:
\begin{equation}
\begin{aligned}
L_3=&\mathbb{E}_{q_\phi(Z,W,H|D)}[ \log p_{\theta}(D|H)] \\
= & \sum_{i,j} \mathbb{E}_{h\sim q_\phi} \log p_{\theta}(c_{ij}|h_{ij}) = \sum_{i,j} \mathbb{E}_{h\sim q_\phi} \Vert c_{ij}-f_{d3}(h_{ij}) \Vert_2^2,
\end{aligned}
\end{equation} 
where $f_{d3}$ is the FNN of decoder 3. For each diffusion induced network $d$, we sample pairs of nodes that are covered by links in $\mcE_d$ (where $e^d_{ij}=1$), and $c_{ij}$ is set to $\mcC_d$.  
During training. we firstly sample a diffusion induced network $d_s$ from $\mcD$, and then only sample positive pairs of nodes \wrt~$\mcE_s$ and make decoder 3 learn to reconstruct the diffusion contents $\mcC_s$ and diffusion structures $\mcE_s$ simultaneously.


For the KL-divergence terms:
\begin{equation} 
\begin{aligned}
&KL(q_\phi(W) \Vert p(W) ) = \sum_{i,j} \sum_{k=1}^K KL(q_\phi(w_k) \Vert \mathcal{N}(0,I))  \\
 = &\sum_{i,j} \sum_{k=1}^K \frac{1}{2} \{\Vert \sigma_k\Vert^2_2  + \Vert \mu_k\Vert ^2_2 -\kappa_H -\log \det(\text{diag}(\sigma_k^2)) )\}.
\end{aligned} 
\end{equation} 
The unit Gaussian is used as the prior for all Gaussian models in $\mcW$. $\kappa_H$ is the dimension of the edge representation $\mcH$.

For edges with no relation labels, we set the prior $p(Z)$ to be the uniform distribution.
When relation labels are available, we set $p(Z)$ to the one-hot distribution and apply Laplace smoothing with parameter $\eta$ to avoid the magnitude explosion of KL-divergence:
\begin{equation}
\begin{aligned}
    & KL(q_\phi(Z|E,A,D)\Vert p(Z))\\
  =& \sum_{i,j, \text{unsup}} \{ \sum_{k=1}^K z_{ijk} \log z_{ijk}  \} + 
   \sum_{i,j, \text{sup}}\{\sum_{k=1}^K z_{ijk} \log\frac{z_{ijk}}{\mathbb{I}(k=z_{ij}^*) + \eta} \},
    \end{aligned}
\end{equation}
where \textit{unsup} and \textit{sup} denote the unsupervised and supervised node pairs respectively, and $z_{ij}^*=k$ means $e_{ij}$ is labeled with the $k$-th relation. Under this setting, the model is trained in a semi-supervised learning fashion, and we only consider single label supervision in this work.

\subsubsection{Training}
Training our model involves the learning of all parameters in the encoder network $q_\phi(Z, W, H|G)$ and decoder network $p_\theta(G|Z,W,H)$. As our multi-modal decoders jointly integrate multiple observed signals on social networks, $p_\theta(G|Z,W,H)$ can be further decomposed into 
\begin{align}
 p_\theta(G|Z,W,H) = p_{\theta_1}(E|H)p_{\theta_2}(A|H)p_{\theta_3}(D|H),
\end{align}
The equation holds because we assume the variable dependence structure in Figure \ref{fig:graphic}, which allows us to learn the whole decoder network $p_\theta$ by iteratively optimizing each of the three decoders \wrt~their corresponding losses. During the iterative training process, each decoder is jointly trained with the same encoder $q_\phi$, which allows the model to effectively integrate the multiple observed signals, capture the underlying relation semantics and regularize it with proper prior knowledge.

The training of each encoder-decoder combination  generally follows that of variational inference for variational autoencoders.
We design an efficient variational inference algorithm with two-step Monte Carlo sampling and reparameterization tricks. It allows joint learning of $\mcW$ and $\mcZ$, together with other non-stochastic parameters in the encoder and decoder networks through principled Bayesian inference. Except for the particular reconstruction losses, the algorithm works in the exact same way for all three decoders. 

\begin{algorithm}[h!]
\small
\caption{\textsc{ReLearn} Training}
\begin{algorithmic}[1]
\Procedure{Training}{}
\Comment{Input}\\
$\mcG$: the social network; $B$: batch size; $T$: number of batches.
\For{$t=1:T$}
\For {$\mcX$ in $\{\mcE, \mcA, \mcD\}$}
\State Sample $B$ pairs of nodes with observed signals of $\mcX$.
\State Use the encoder network to compute $q_\phi(Z|G)$.
\For {$k=1:K$}
\State Draw $B$ random variables $\epsilon_k \sim \mcN(0, I)$.
\State Compute $\hat{W}_k=\mu_k+\sigma_k\epsilon_k$. 
\State Draw $B$ random variables $G_k \sim$ Gumbel(0, 1).
\State Compute $\hat{Z}_k=\frac{\exp((\log(Z_k)+G_k)/\tau)}{\sum_{k'=1}^K\exp((\log(Z_{k'})+G_{k'})/\tau)}$.
\EndFor
\State Compute $H=\sum_{k=1}^K \hat{Z}_k\odot\hat{W}_k$.
\State Use the decoder network to compute $p_\theta(X|H)$.
\State Compute the ELBO with $q_\phi$ and $p_\theta$.
\State Update $\{\phi, \theta\}$ with gradient backpropagation.
\EndFor
\EndFor
\EndProcedure
\end{algorithmic}
\end{algorithm}

Without loss of generality, in Algorithm 1, we again use $\mcX$ to refer to any of the three signals to describe our training process. 
In Line 8-9 and 10-11, we apply the reparameterization trick to $\mcW$ and $\mcZ$ by drawing random samples from the standard Normal distribution and Gumbel distribution \cite{jang2017categorical, maddison2017concrete}, respectively, which allows us to push the randomness to the continuous variables $\epsilon$ and discrete variables $G$, and directly optimize the encoder parameters $\phi$ through standard backpropagation. 

As shown in Algorithm 1, besides the sampling process which takes $O(1)$ time for each batch, the whole training process of \textsc{ReLearn} can be done through standard stochastic gradient backpropagation, which allows us to fully leverage well-developed optimization software like mini-batch adam \cite{kingma2014adam} and hardware like GPU. Due to the inductive nature of \textsc{ReLearn}, we do not need to enumerate every pair of nodes in the network. Therefore, the overall computational complexity of training is $O(TBK)$, which are all constant numbers irrelevant of the network size. In other words, the actual training time of \textsc{ReLearn} depends more on the quality and consistency of the network signals than the size of the network.

In our experiments, we observe that the training of \textsc{ReLearn} often converges with $TB=\rho|V|$ with $\rho \in [1, 10]$, which gives a rough computational complexity of $O(|V|)$, where $|V|$ is the number of nodes. This often leads to much less training time than most baselines on the same networks.


\section {Experiments}
\label{sec:exp}

\subsection{Experimental Settings}
\subsubsection{Datasets}
We use two public DBLP author networks and two internal LinkedIn member networks for our experiments. 

In the DBLP networks, nodes are authors and links are co-authorships. Node attributes are generated from publications and information diffusions are generated from citations. 
Particularly, user attributes are computed by averaging the word embedding\footnote{http://nlp.stanford.edu/data/glove.840B.300d.zip} of keywords and titles in their publications, which are 300-dim.
Information diffusions are generated by firstly selecting papers with $10-100$ citations, and construct author subnetworks by including authors who cite the corresponding papers and their links. Diffusion contents are then the paper embedding of the cited paper, which are also 300-dim.
We use the ground-truth relation labels of \textsf{advisor-advisee} and \textsf{colleague} relations from \cite{wang2010mining}. A subnetwork DBLP-Sub is generated by including all pairs of authors with ground-truth relation labels and their direct co-authors. DBLP-All is the whole network with all authors and links on DBLP.

In the LinkedIn networks, nodes are members (users) and links are bi-directional member connections. We generate two relatively small and complete networks of members in \textsf{Bay Area, US} and \textsf{Australia}. 
Node attributes are generated based on the anonymous user profiles, including features like \textsf{skills}, \textsf{locations}, \textsf{languages} and so on. Numerical features like \textsf{longitudes} and \textsf{latitudes} are directly adopted, whereas categorical features like \textsf{skills} and \textsf{languages} are firstly converted into bag-of-skill and bag-of-language vectors, and then further reduced to smaller dimensions via incremental PCA\footnote{https://scikit-learn.org/stable/auto\_examples/decomposition/plot\_incremental\_pca.html}. The final dimension of user attributes is $466$. 

Ideally, information diffusions should be generated based on public posts, such as \textsf{popular articles} shared by users. However, due to privacy concern, we could not get that data in this work. Alternatively, we use users' following of influential individuals to model the influence propagation. This following relation is one-directional and different from connections, which we believe to be indicative to users' personal interests. Particularly, we randomly choose influential individuals with $10-100$ followers and generate diffusion induced networks by including the followees and their own connections. Diffusion contents are generated by embedding the textual descriptions of the influential individuals from their profile, by averaging the word embedding in the same way as we do for papers on DBLP. The diffusion content vectors are 300-dim.

To generate the ground-truth relation labels, if two connected members attend the same school in the same time, we label their relation as \textsf{schoolmate}, and the same is done for \textsf{colleague}.
Note that, we exclude the \textsf{education} and \textsf{working experience} for generating node attributes, because they are highly correlated with the ground-truth relation we use for evaluation. However, this does not weaken the utility of our model, since this reliable generation of \textsf{schoolmate} and \textsf{colleague} relations can only cover a small portion of all observable connections ($<0.3\%$). Moreover, \textsc{ReLearn} can be used to learn many other relations that cannot be easily verified or even defined (\eg, relatives, townsmen, close friends), in an unsupervised way.


\begin{table}[h!]
\centering
\small
 \begin{tabular}{|c||c|c|c|c|c|c|}
   \hline
{\bf Dataset}&{\bf \#Nodes}&{\bf \#Links}&{\bf \#Diff.}&{\bf Rel.(\%)}\\
  \hline
  \hline
{\bf DBLP-Sub}& 23,418 & 282,146 & 100,859 & 0.4341 \\
\hline
{\bf DBLP-All}& 1,476,370 & 4,109,102  & 410,822 & 0.0196 \\
\hline
{\bf LinkedIn-Bay}& 1,481,521 & 67,819,313 & 45,686 & 0.2239 \\
\hline
{\bf LinkedIn-Aus}& 6,598,127 & 328,005,877 & 129,510 & 0.1592  \\
\hline
 \end{tabular}
 \caption{ \label{tab:stats}Statistics of the four datasets we use. $\#$Diff$.$ is the number of information diffusions, and Rel$.(\%)$ is the coverage of labeled relations over all observable links.}
\end{table}

\subsubsection{Compared algorithms} 
Since the problem setting of \textsc{ReLearn} is quite different from relation learning on knowledge graphs, we find a comprehensive list of baselines from the state-of-the-art on network inference and embedding. However, none of the existing models can combine all signals as we consider in this work. Besides existing baselines, we also compare multiple variants of \textsc{ReLearn} to provide in-depth understanding over the utilities of different model components.
\begin{itemize}[leftmargin=10pt]
\item \textbf{GraphSage} \cite{hamilton2017inductive}: One of the strongest and most efficient variants of the popular GCN model that integrates node attributes and link structures for learning network embeddings. 
\item \textbf{STNE} \cite{liu2018content}: The state-of-the-art unsupervised text-rich network embedding algorithm based on self-translation of sequences of text embeddings into sequences of node embeddings.
\item \textbf{PTE} \cite{tang2015pte}: Extension of the popular network embedding algorithm LINE \cite{tang2015line} into text-rich network embedding. We also enable supervision for PTE by constructing multiple bipartite graphs connected by links with different relation labels.
\item \textbf{Planetoid} \cite{yang2016revisiting}: Extension of the popular network embedding algorithm DeepWalk \cite{perozzi2014deepwalk} into text-rich network embedding. We also enable supervision for Planetoid through pair-wise sampling for relation prediction.
\item \textbf{TopoLSTM} \cite{wang2017topological}: One of the state-of-the-art diffusion prediction model with network embedding. Embedding of edges not covered by any diffusion is computed as the average of the embedding of all neighboring edges.
\item \textbf{Inf2vec} \cite{feng2018inf2vec}: Another State-of-the-art diffusion prediction model with network embedding. The same process for TopoLSTM is done for edges not covered by any diffusion.
\item \textsc{ReLearn w/o diff}: To study the ability of \textsc{ReLearn} in integrating multiple signals, we decompose the model by removing each decoder. As an example, we show the performance of \textsc{ReLearn} \textit{without} decoder 3 (the information diffusion decoder). We find that with the additional attribute decoder, this model variant still performs better than the base model of GVAE \cite{kipf2016variational}.
\item \textsc{ReLearn w/o vae}: To study the effectiveness of our novel Gaussian mixture VAE in capturing the latent relations, we remove VAE and directly use the output of the graph edge encoder as the edge representation and input of the multi-modal decoders.
\item \textsc{ReLearn w/o sup}: The unsupervised training version of \textsc{ReLearn} by using the uniform multinomial distribution as the prior for the mixture weights $Z$ for all edges. 
\item \textsc{Relearn}: Our full \textsc{ReLearn} model\footnote{Code available at: https://github.com/yangji9181/RELEARN}.
\end{itemize}
The implementations of all existing baselines are provided by their original authors and the parameters are either set as the default values or tuned to the best via standard five-fold cross validation. 
As for \textsc{ReLearn}, for the encoder network, we use a two-layer GCN, with embedding sizes $200$ and $100$. We set the number of sampled neighbors to $30$. After that, we use a single-layer FNN of size $3$ with ReLU activations. The edge embedding size and dimension of Gaussian mixtures are set to 100. For the decoder network, we use three $2$-layer FNNs with ReLU activations for the three signals, with sizes $200$ and $300$. For link reconstruction, we set the positive-negative sampling ratio to $1$. The weights of three decoders are simply set to the same. The number of latent relations are set to 2 for all datasets.
For training, we set the batch size to $1024$ and learning rate set to $0.001$ on all datasets. For DBLP datasets, we set the number of batches to $500$, and for LinkedIn datasets, we set the number of batches to $5000$.

\subsubsection{Evaluation metrics}
The node embeddings learned by all compared algorithms are concatenated into edge embeddings and then fed into MLPs with the same structure, which is then trained and tested on the same splits of labeled relations.
Standard classification accuracy is computed based on the prediction of the MLPs using the network embedding generated by different algorithms. 
To observe significant differences in performance, we run all algorithms on 5 different training-testing splits of relation labels to record the means and standard deviations. Then we conduct paired statistical t-tests by putting \textsc{ReLearn} against all baselines.

\subsection{Performance Comparison with Baselines}
We quantitatively evaluate \textsc{ReLearn} against all baselines on the task of relation learning.
Table \ref{tab:perform} shows the classification accuracy evaluated for all compared algorithms. 
The results all passed the significant t-tests with $p$-value 0.01.

As we can see in Table \ref{tab:perform}, \textsc{ReLearn} constantly outperforms all baselines by significant margins on all datasets, while the compared algorithms have varying performances. 
Taking a closer look at the results on different datasets, we observe that the task of learning the \textsf{schoolmate} and \textsf{colleague} relations on LinkedIn is much harder than the \textsf{adviser-advisee} and \textsf{colleague} relations on DBLP. This is probably because the social contents and links are often more noisy and complex than those in the publication networks. \textsc{ReLearn} excels on both of the LinkedIn networks, outperforming the best baseline by 17.9\% and 28.5\%, respectively. Such significant improvements strongly indicate the power of \textsc{ReLearn} in capturing complex noisy signals on social networks for high-quality relation learning.
Moreover, the full \textsc{ReLearn} model also consistently outperforms all other \textsc{ReLearn} variants, which further corroborates the effectiveness of \textsc{ReLearn} in integrating multi-modal network signals and limited supervision.

%
%

\begin{table*}[]
 \centering
 \small
 \begin{tabular}{|c|c|c|c|c|}
 \hline
Algorithm & DBLP-Sub & DBLP-All & LinkedIn-Bay & LinkedIn-Aus \\
\hline
GraphSage & $0.8596\pm0.0201$ & $0.8482\pm0.0158$ & $0.6139\pm0.0367$ & $0.5831\pm0.0072$\\
\hline
STNE & $0.7577\pm0.0425$ & $0.7434\pm0.0214$  & $0.5695\pm0.0236$ & $0.5554\pm0.0160$\\
\hline
PTE & $0.7265\pm0.0018$ & $0.6988\pm0.0222$ & $0.5636\pm0.0378$ & $0.5549\pm0.0041$ \\
\hline
Planetoid & $0.8531\pm0.0205$ & $0.8686\pm0.0206$ & $0.5608\pm0.0301$  & $0.5448\pm0.0045$ \\
\hline
TopoLSTM & $0.6675\pm0.0435$ & $0.7374\pm0.0149$ & $0.5874\pm0.0257$  & $0.5616\pm0.0062$ \\
\hline
Inf2vec & $0.6618\pm0.0401$ & $0.7453\pm0.0181$ & $0.6198\pm0.0388$ & $0.5848\pm0.0068$ \\
\hline
\hline
\textsc{ReLearn w/o diff} & $0.8890\pm0.0031$ & $0.8465\pm0.0138$ & $0.6616\pm0.0390$ & $0.6934\pm0.0022$ \\
\hline
\textsc{ReLearn w/o vae} & $0.8433\pm0.0154$ & $0.8376\pm0.0060$ & $0.6293\pm0.0194$ & $0.6626\pm0.0087$ \\
\hline
\textsc{ReLearn w/o sup} & $0.8947\pm0.0170$ & $0.8980\pm0.0115$ & $0.6771\pm0.0211$ & $0.7134\pm0.0048$ \\
\hline
\textsc{ReLearn} & $\bf{0.9224\pm0.0026}$ & $\bf{0.9208\pm0.0042}$ & $\bf{0.7308\pm0.0457}$ & $\bf{0.7514\pm0.0033}$ \\
\hline
 \end{tabular}
 \caption{\label{tab:perform}\textbf{Relation learning accuracy of compared algorithms on four real-world social networks.}}
\end{table*}

\begin{table*}[h!]
 \centering
 \scriptsize
 \begin{tabular}{|l|l|l|}
 \hline
Area \rom{1} & Area \rom{2} & Area \rom{3} \\
\hline
Define $|$ Create $\ldots$ Implement $|$ Support $|$ Succeed & Writer, Dancer, Entrepreneur $\ldots$ & $\ldots$ Benefits Negotiation, Salary Negotiation \\
Training, Program Development, Exercise Prescription $\ldots$ & Blogger \& Youtuber $\ldots$ & Corporate Advisor $|$ Investment Banker $\ldots$ Shareholder Representative \\
$\ldots$ Sponsorship Program Development, Fellowship Application & FASHION, BEAUTY, TRAVEL, LIFE $\ldots$ & Project Manager $|$ Leader $\ldots$ Performance Manager $|$ PA/EA  \\
$\ldots$ Talent Management \& Success Planning & Social \& Environmental Justice $\ldots$ & Recruitment, Performance Management $\ldots$ Gap Management \\
Talent Acquisition, Recruiting, Head Hunting $\ldots$ & Chef Traditional Italian $\ldots$ Proactive & Change \& Transition Management, Programme Management $\ldots$ \\
Recruitment $\ldots$ Development, Relationship Management & Wellness Coach-Clean Food $\ldots$ Warrior-Positive Thinker & People Management, Performance Coaching, Human Resource $\ldots$ \\
An Entrepreneur. A Scholar $\ldots$ & Food $\ldots$ Driven \& Hungry & $\ldots$ Beautiful Web Design \& Digital Media Solutions \\
Portfolio Building $|$ Training $\ldots$ & $\ldots$ A Bohemian Fashion Boutique & Test Automation, Test Management, Technical Testing $\ldots$ \\
Learning \& Development, Organisational Culture, Engagement $\ldots$ & Licensed Waterproofing Technician $\ldots$ & Intellectual Property $\ldots$ \\ 
\hline
 \end{tabular}
 \caption{\label{tab:case}\textbf{Decoded diffusion contents on edges generated with three different latent relations.}}
\end{table*}

\subsection{In-depth Model Analysis}
To comprehensively evaluate the performance of \textsc{ReLearn} in comparison with the baselines, we design a series of in-depth analysis, by varying the amount of training data, as well as adding noise and sparsity to the network signals. 

\begin{figure}[h!]
\centering
\hspace{-8pt}
\subfigure{
\includegraphics[width=0.238\textwidth]{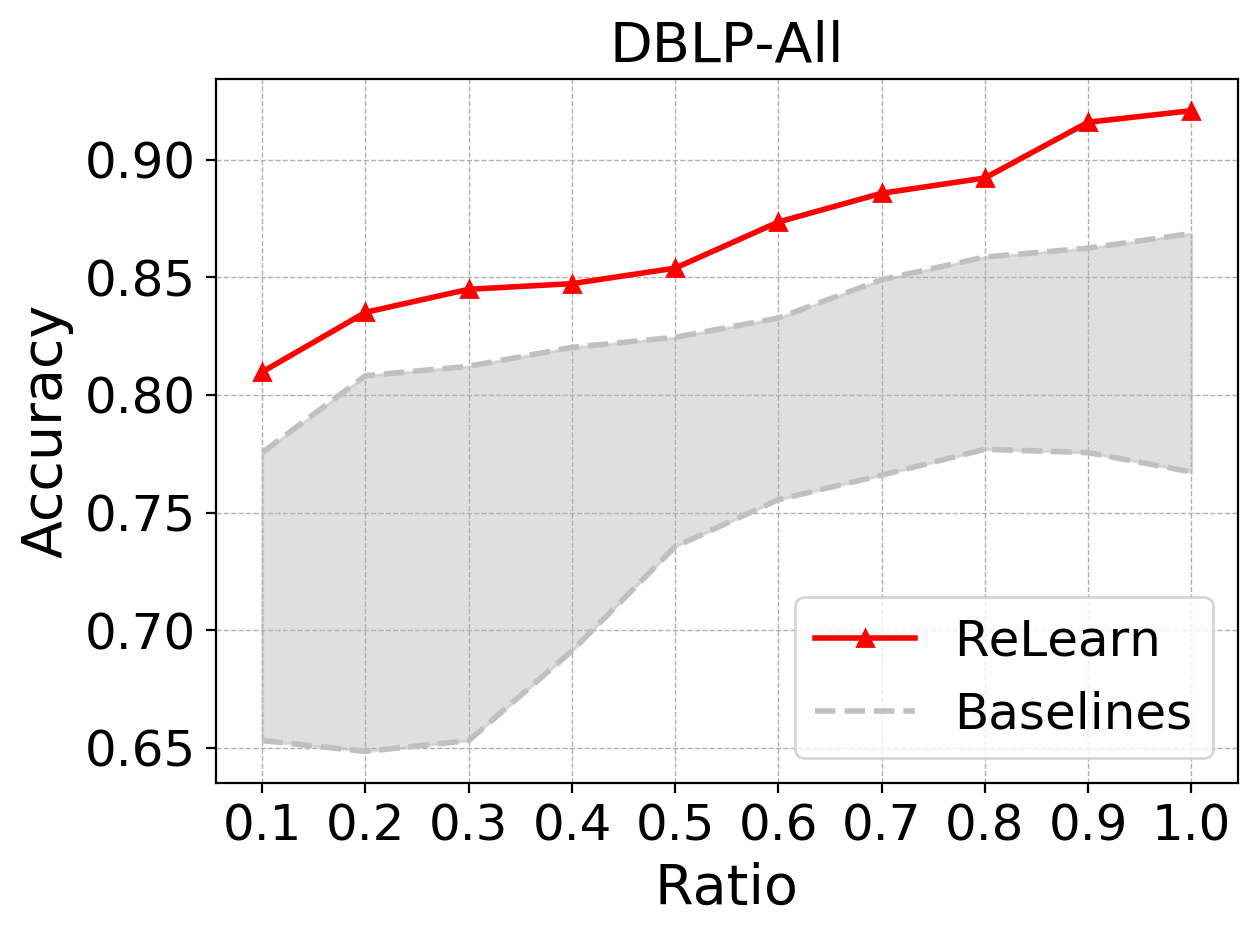}}
\hspace{-8pt}
\subfigure{
\includegraphics[width=0.238\textwidth]{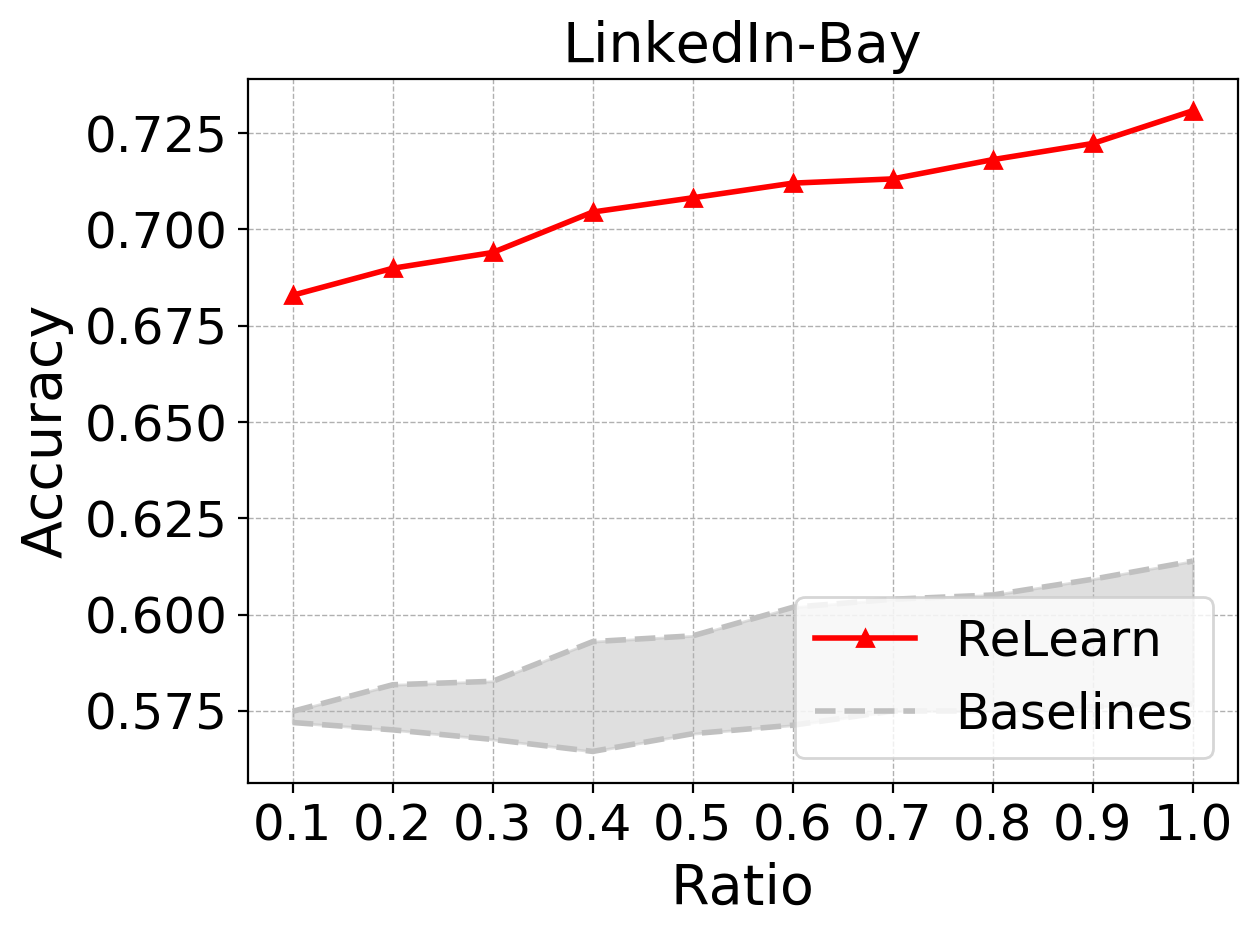}}
\hspace{-8pt}
\caption{\textbf{Varying amounts of training data.}}
\label{fig:label}
\end{figure}

\header{Efficiency towards limited training data.}
One major challenge of relation learning on social networks is the lack of high-quality relation labels. Therefore, an ideal model should be efficient in leveraging limited training data. To study such efficiency of \textsc{ReLearn}, we conduct experiments on all datasets with varying amounts of training data. Particularly, for each of the 4:1 splitting of training and testing data, we use 10\% - 100\% of the 80\% training data to train \textsc{ReLearn} and all compared algorithms, and evaluate on the 20\% testing data. The results on DBLP-All and LinkedIn-Bay are presented in Figure \ref{fig:label}.

\begin{figure}[h!]
\centering
\hspace{-8pt}
\subfigure{
\includegraphics[width=0.238\textwidth]{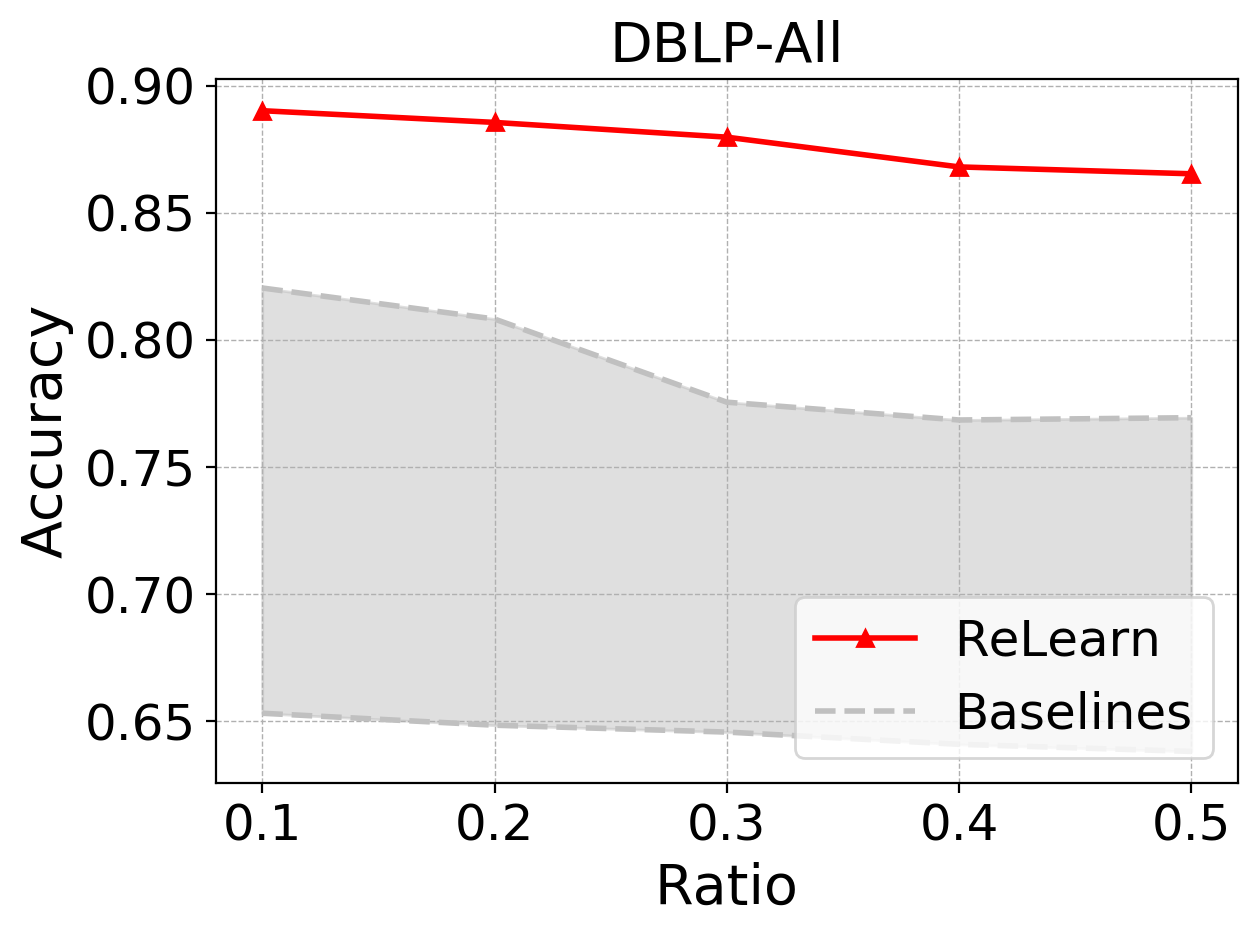}}
\hspace{-8pt}
\subfigure{
\includegraphics[width=0.238\textwidth]{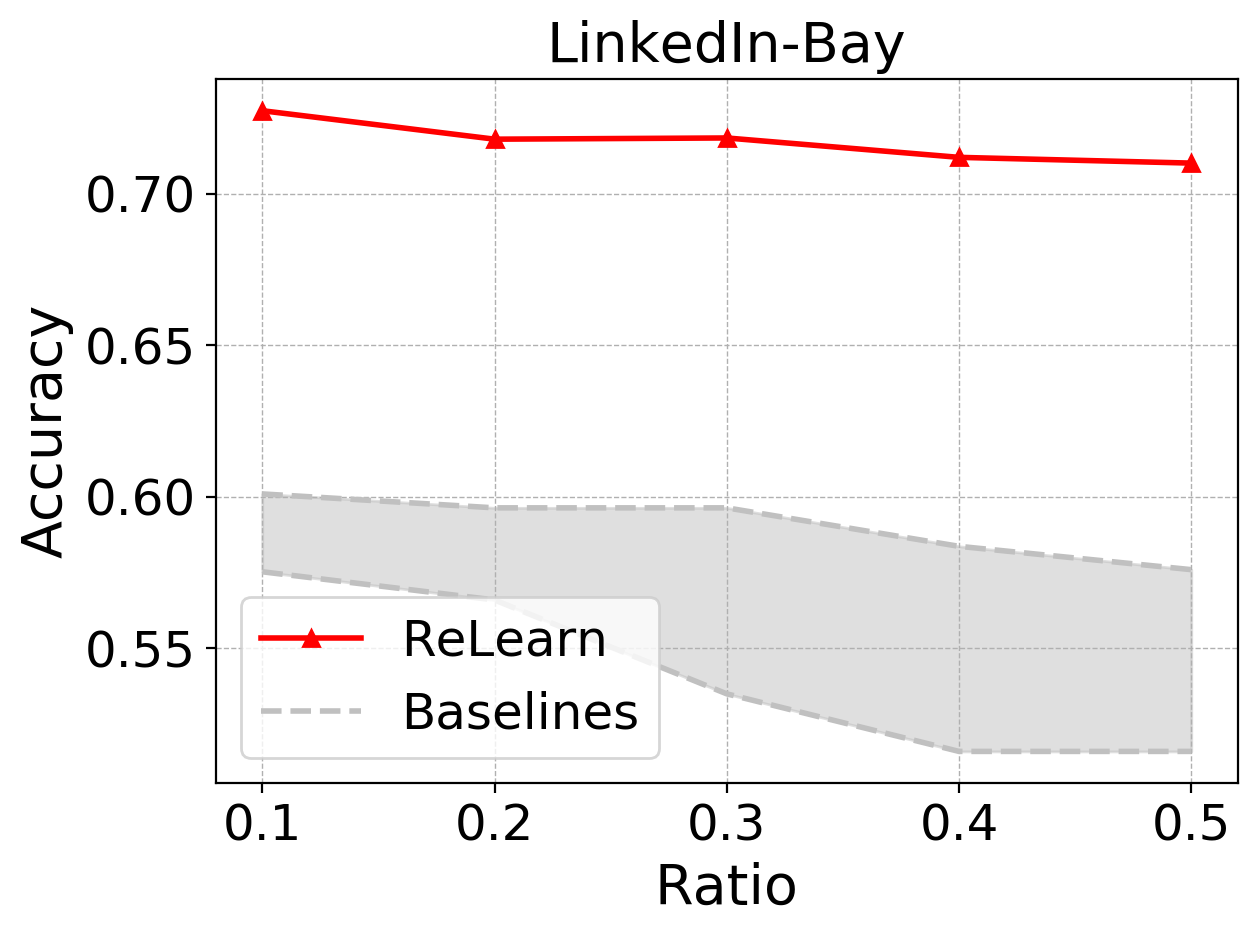}}
\hspace{-8pt}
\caption{\textbf{Varying amounts of attribute noise.}}
\label{fig:noise}
\end{figure}

\header{Robustness towards attribute noise.}
On real-world social networks, user attributes are often highly noisy, since users might fill in various free-style contents and even random contents. Therefore, an ideal model for relation learning should be robust towards attribute noise. To study such robustness of \textsc{ReLearn}, we conduct experiments on all datasets by adding different amounts of random noise onto the user attributes. Particularly, since all models take the normalized numerical embedding of attributes as input, we add the unit multivariate Gaussian noise scaled by 0.1-0.5 to the attribute vector of each user. The modified input for all compared algorithms (including \textsc{ReLearn}) is the same. The results on DBLP-All and LinkedIn-Bay are presented in Figure \ref{fig:noise}.

\begin{figure}[h!]
\centering
\hspace{-8pt}
\subfigure{
\includegraphics[width=0.238\textwidth]{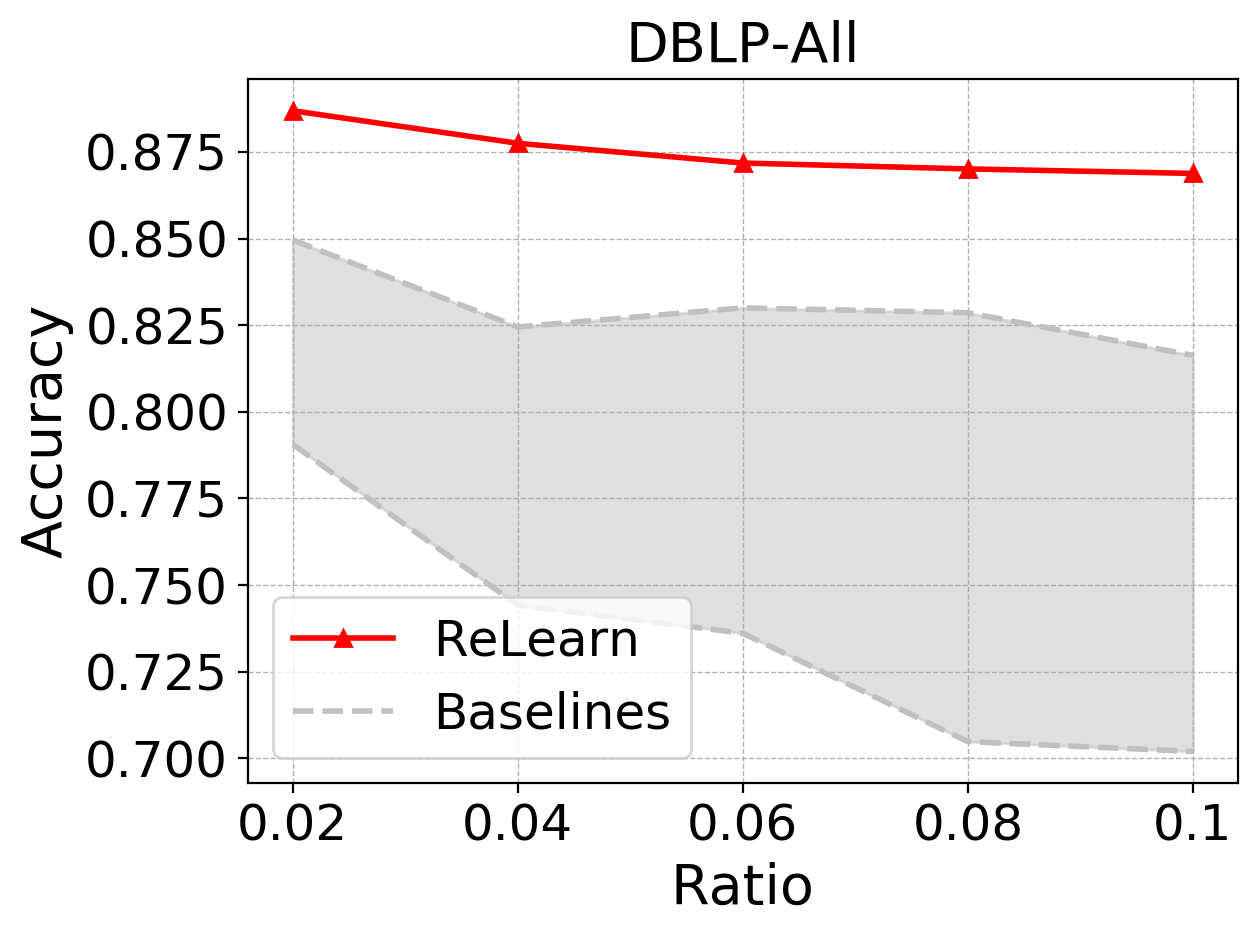}}
\hspace{-8pt}
\subfigure{
\includegraphics[width=0.238\textwidth]{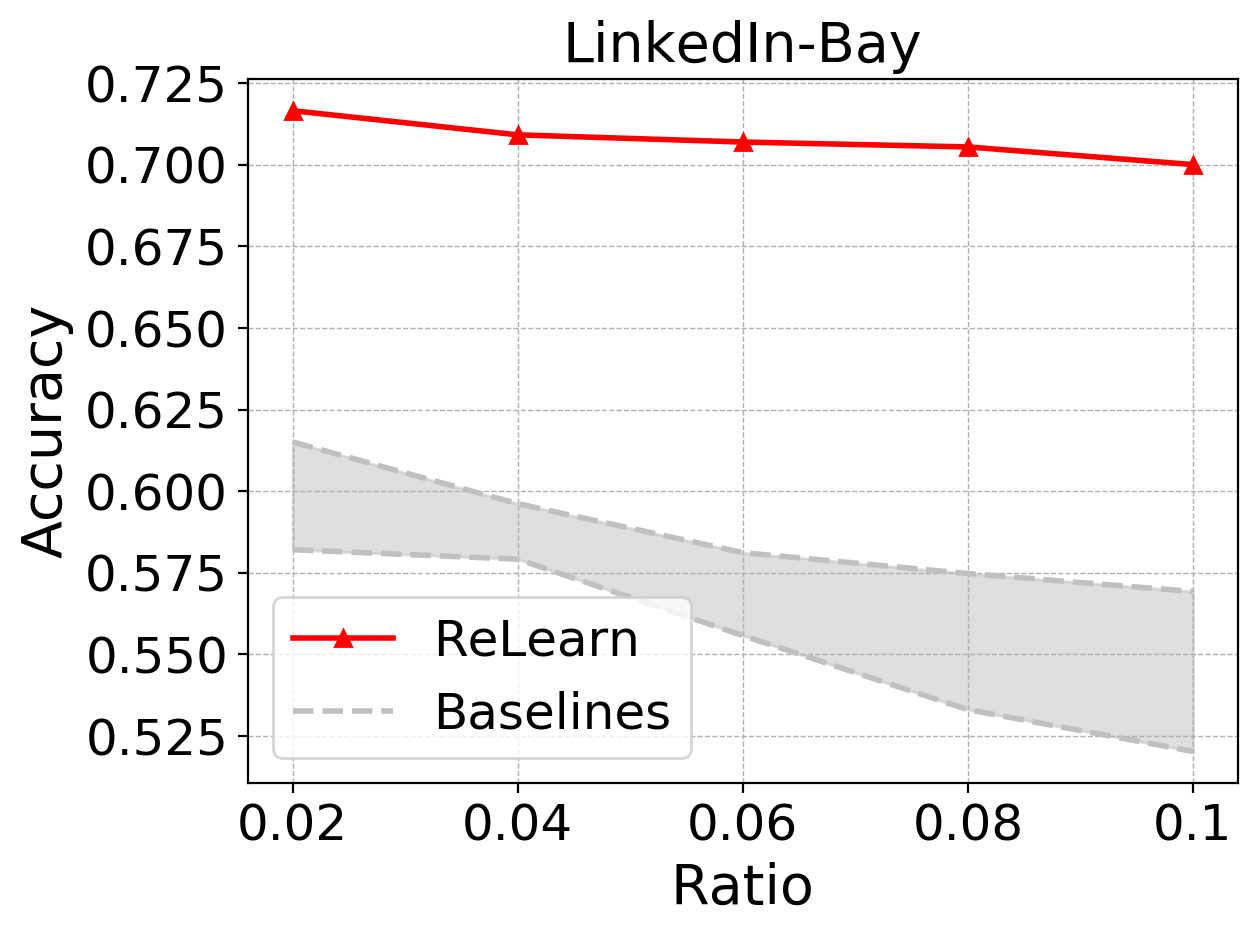}}
\hspace{-8pt}
\vspace{-5pt}
\caption{\textbf{Varying amounts of link removal.}}
\vspace{-5pt}
\label{fig:link}
\end{figure}

\header{Robustness towards missing links.}
On real-world social networks, real-world friends may not necessarily have established links. Therefore, an ideal model for relation learning should be robust towards missing links. To study such robustness of \textsc{ReLearn}, we conduct experiments on all datasets by randomly removing existing links in the network. Particularly, we randomly remove 2\%-10\% of links in the whole networks. The modified input for all compared algorithms (including \textsc{ReLearn}) is the same. The results on DBLP-All and LinkedIn-Bay are presented in Figure \ref{fig:link}.

\header{Remarks on runtimes.}
While the exact runtimes of compared algorithms are hard to determine due to different convergence rates of each train, we observe that the runtime of \textsc{ReLearn} is close to the more efficient baselines like PTE, Planetoid and GraphSage, and is often significantly shorter than the heavier baselines like STNE, TopoLSTM and Inf2vec. 

\section {Case Studies}
\label{sec:case}
\begin{figure}[h!]
	\centering
	\includegraphics[width=0.8\linewidth]{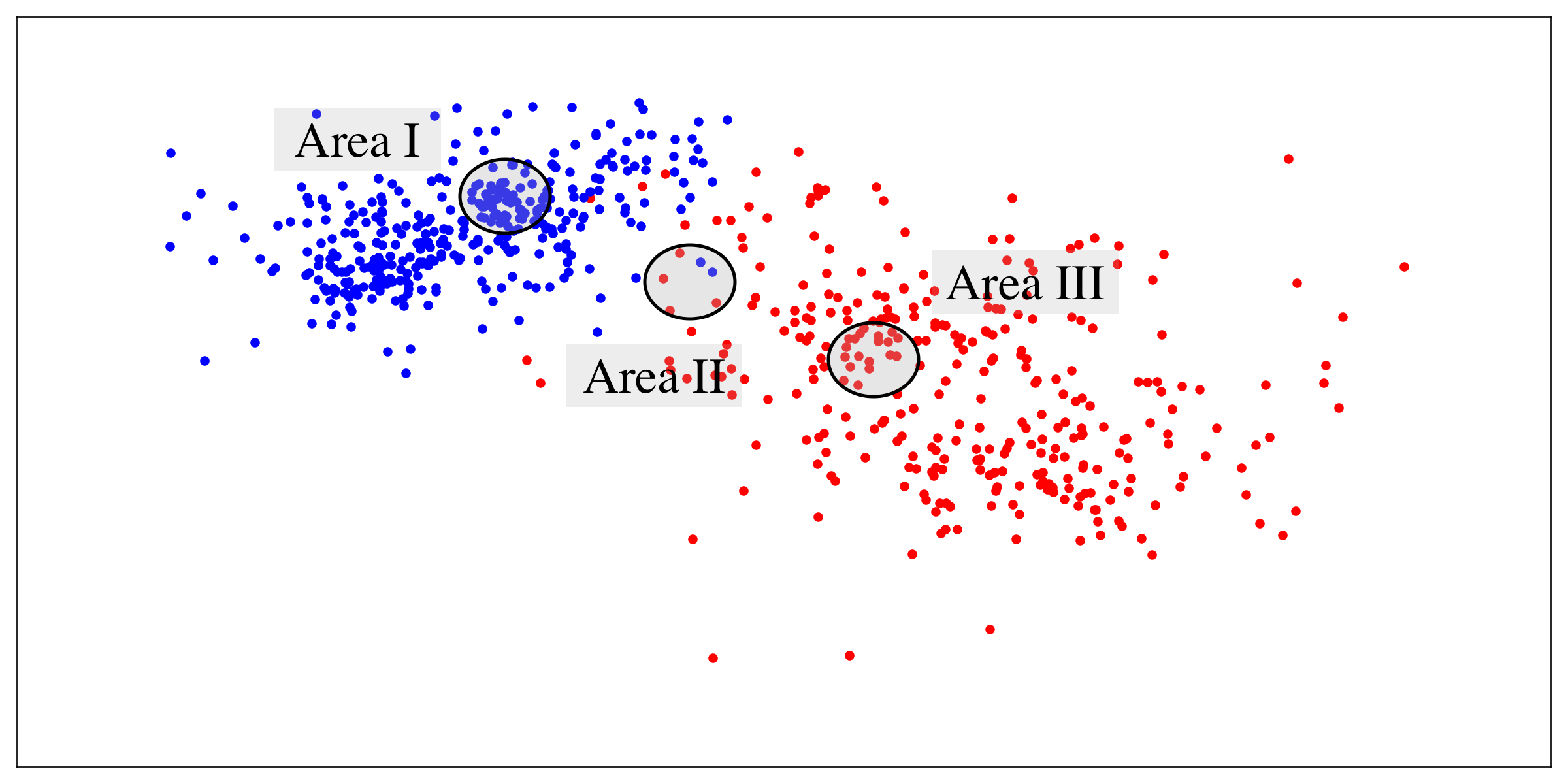}
	\caption{Visualization of edge representations on LinkedIn-Aus computed by \textsc{ReLearn}. Red and blue colors denote the ground-truth labels of \textsf{schoolmate} and \textsf{colleague}.}
	\label{fig:vis}
	\vspace{-8pt}
\end{figure}

To observe how \textsc{ReLearn} captures the relation semantics among users with learned edge representations, we visualize the embedding space by plotting some of the labeled edges in the LinkedIn-Aus network. We employ standard PCA to reduce the embeddings from 100-dim to 2-dim for plotting. As we can see from Figure \ref{fig:vis}, edges carrying the two relations clearly form two clusters. 

Moreover, the generative nature of \textsc{ReLearn} allows us to further interpret the learned latent relations, by sampling edge representations from the learned Gaussian mixture model and decoding them with the multiple learned decoders. This is especially useful in the unsupervised learning scenario, where besides the latent distributions, we also want to make sense of the learned relations. 

In Table \ref{tab:case}, as an example, we show the decoded textual feature from decoder 3 (\ie, the information diffusion decoder), which provides valuable insights into the learned relations. The edge representations are generated by sampling from the Gaussian distribution of $W_1$, $W_2$ and a uniform mixture of $W_1$ and $W_2$, which roughly corresponds to the three marked areas in Figure \ref{fig:vis}.

As we can observe in Table \ref{tab:case}, edges in Area \rom{1} likely carry the \textsf{schoolmate} relation, with decoded contents mainly about \textsf{Learning} and \textsf{Advising}, whereas Area \rom{3} clearly corresponds to \textsf{colleagues}, due to decoded topics like \textsf{Management} and \textsf{Performance}. Edges in Area \rom{2} hold a mixture of the two relations, with more personal life oriented contents like \textsf{Food}, \textsf{Travel}, \textsf{Wellness}, \etc. Although the encoder does not directly consider information diffusion, it effectively helps the decoder to capture this information during the joint training process.

Note that, in this example, we already know that the two relations we learn are \textsf{schoolmates} and \textsf{colleagues}, which we use as a verification of the utility of \textsc{ReLearn}. In the more realistic situations where we have no access to ground truth, the multiple decoders of \textsc{ReLearn} still provide meaningful interpretations over the learned relations, which are valuable for downstream services like relation-specific friendship recommendation and content routing.

\section{Conclusion}
In this work, for the novel and challenging problem of relation learning on social networks, we develop \textsc{ReLearn}, a multi-modal graph edge variational autoencoder framework to coherently combine multiple signals on social networks towards the capturing of underlying relation semantics on user links. Moreover, the generative nature of \textsc{ReLearn} allows us to sample relational pairs for interpreting the learned relations, while its inductive nature enables efficient training regardless of the network sizes. Finally, the general and flexible design of \textsc{ReLearn} makes it readily applicable to any real-world social platforms with multi-modal network signals, where the learned node and edge embeddings can be used to improve the targeting of various downstream services.
\bibliographystyle{ACM-Reference-Format}
\bibliography{carlyang} 

\end{document}